# A search for leptonic photon $Z_l$ at all three CLIC energy stages by using artificial neural networks (ANN)

S. O. Kara[1,*](Corresponding Author, no funding), S. Akkoyun[2,**] (no funding)

[1] Niğde Ömer Halisdemir University, Bor Vocational School, 51240, Niğde, Turkiye

[2] Cumhuriyet University, Faculty of Sciences, Department of Physics, 58140, Sivas, Turkiye

*Abstract:* In this work, the possible dynamics of the massive leptonic photon $Z_l$ are reconsidered via the process $e^+e^- \to \mu^+\mu^-$ at Compact Linear Collider (CLIC) with updated center of mass energies ($380\ GeV, 1500\ GeV\ and\ 3000\ GeV$). We show that the new generation colliders as CLIC can observe massive leptophilic vector boson $Z_l$ with mass up to the center of mass energy, provided that leptonic coupling constant is $g_l \geq 10^{-3}$. In this study, we also estimated the cross-sections by artificial neural networks using the theoretical results we obtained for CLIC. According to the results obtained, it was seen that these predictions could be made through machine learning.

*e-mail: sokara@ohu.edu.tr

**e-mail: sakkoyun@cumhuriyet.edu.tr

## 1. Introduction

The idea of massless leptonic vector particle was first proposed by L. B. Okun [1]. Experiments testing the equality of inertial and gravitational masses [2] have put a very strong limit for the interaction constant of the massless leptonic photon with matter: $\alpha_l < 10^{-49}$. When this constant is compared to the electromagnetic interaction constant ($\alpha_{EM} \approx 10^{-2}$), it is natural to conclude that massless leptonic photons coupled to the lepton charge do not exist. However, the use of an extra range for the leptonic interaction constant [3] has revived interest in this subject and led to many publications[4-11].

In [12], it was phenomenologically shown that lepton charge coupled to leptophilic massive vector boson and the possible dynamics associated with these charges in future linear colliders as International Linear Collider (ILC) and Compact Linear Collider (CLIC). The CLIC is optimized to be built and operated at collision energies of $380\ GeV, 1500\ GeV\ and\ 3000\ GeV$ [13]. Many works in the literature have assessed CLIC's reach to vector bosons in different contexts [14-20].

In this paper we have reconsidered phenomenology of massive leptonic photon, $Z_l$, at all three CLIC energy stages, using the updated key parameters as in [21]. We also estimated the cross-sections by artificial neural networks using the theoretical results we obtained for CLIC. Recently, ANN method has been applied for the determination of cross-sections in high-energy physics. Addepalli has been applied machine learning techniques for cross-section measurements for the vector-boson fusion production of the Higgs boson [22]. Mekosh used machine learning to search for vector boson scattering at the CMS detector [23]. Sauerburger performed H→ττ cross section measurements using machine learning in the ATLAS detector [24]. Akkoyun and Kara did an approximation to the cross sections of $Z_l$ boson production at CLIC by using neural networks [25]. Kara et al. used neural network method in order to investigate leptophilic gauge boson $Z_l$ at ILC [26]. In the present study, we have used ANN to prediction of cross-sections. According to the results obtained, it was seen that these predictions could be made through machine learning. In section 2, the model is formulated and artificial neural network (ANN) has been explained. Production of the $Z_l$ boson at the CLIC energy stages is analyzed in section 3. In section 4, the ANN results is presented. The results obtained are summarized in the final section.

## 2. Material and Methods

### 2.1 The Model

In our model, to gauge the leptonic quantum number a new $U'_l(1)$ global abelian symmetry is added to standard model (SM) gauge group $(SU_C(3) \times SU_W(2) \times U_Y(1))$. With the experimental discovery of neutrino oscillations [27], the idea of conserving electron, muon and tau lepton charges individually is invalidated. Therefore, we consider single lepton charge which is the same for e, μ, τ and corresponding neutrinos. Also for the interaction of the electroweak vector bosons with fermions, we introduce the following replacement in the free fields Lagrangian:

$$\partial_\mu \to D_\mu = \partial_\mu - ig_2 \mathbf{T} \cdot \mathbf{A}_\mu - ig_1 \frac{Y}{2} B_\mu - ig_l a_l B'_\mu \qquad (2.1)$$

where $g_2$, $g_1$ and $g_l$ are interaction constants, $\mathbf{T}$ is an isospin operator of a corresponding multiplet of fermionic or Higgs fields, $Y$ is hypercharge and $a_l$ is lepton charge of the corresponding multiplet, $\mathbf{A}_\mu$, $B_\mu$, $B'_\mu$ are gauge fields. A mechanism is needed to provide mass to the leptophilic $Z_l$ vector boson that coincides with the $B'_\mu$ vector field. Therefore, Higgs field with lepton charge must be added to the model. Interaction Lagrangian, obeying the $SU_C(3) \times SU_W(2) \times U_Y(1) \times U'_l(1)$ gauge symmetry, can be decomposed as:

$$L = L_{SM} + L' \qquad (2.2)$$

where $L_{SM}$ is standard model Lagrangian and $L'$ is given by:

$$L' = \frac{1}{4} F'_{\mu\nu} F^{\mu\nu\prime} + g_l J^\mu_{lep} B'_\mu + (D_\mu \Phi)^\dagger (D_\mu \Phi) + \mu^2 |\Phi|^2 - \lambda |\Phi|^4 \qquad (2.3)$$

where

$$F'_{\mu\nu} = \partial_\mu B'_\nu - \partial_\nu B'_\mu \qquad (2.4)$$

is field strength tensor,

$$J^\mu_{lep} = \sum_l a_l [\bar{\nu}_l \gamma^\mu \nu_l + \bar{l} \gamma^\mu l] \qquad (2.5)$$

is leptonic current interacting with leptopfilic $Z_l$, $\Phi$ is singlet complex scalar Higgs field.

### 2.2 Artificial Neural Network (ANN)

The ANN method is one of the alternative strong tools for the physics problems [28]. It mimics the brain functionality and nervous system. ANN, which is the base of artificial intelligence, can learn the structure of the data and relationship between them using appropriate algorithms. ANN is composed of mainly three different layers which are input, hidden and output layers. Each layer includes its own neurons. The data is taken from the outside by the input layer neurons as inputs and the output data is the desired one which is exported from output layer neurons. The number of input neurons depends on the problem. As clearly known, inputs are the independent variables of the problem, outputs are the dependent ones.

The number of output neurons is the number of output variables of the problem. Input and output layers, there is one or more layer in which data is mainly processed in this layer. This layer is called as hidden layer and it is crucial to solve non-linear problems. There is no common rule for the determination of the numbers of hidden layer and the hidden neuron. These numbers are independent of the problem. In the fully connected feed-forward ANN model which was used in the present study, data flows in one direction from input to outputs neurons. Each neuron in the layers is connected to all other neurons in the next layer. Therefore, all hidden and output layer neurons have at least one entry. All the entries to the neurons are multiplied by the weight values of their connections and then summed to get the net inputs of the neurons. After obtaining net input to the neuron, it is activated by an appropriate activation function and the outcome is generated. This information is transmitted to the neurons in the next layer by weighted connections. In the case of the output neurons, the outcome is the solution of the problem. In the ANN calculations, all data belonging to the given problem has been divided into two main separate sets. The first part of the data (75% in the present study) is used for the training of ANN. To see the generalization ability of the method, it must be tested over another set of data which is test dataset (25% in the present study). The main task in the training is determination of the values of weighted connections between neurons. In other words, in the training process, it is aimed to find the best weight values which give the best estimation starting from the input. Therefore, the weight values are modified until the acceptable error level between desired and neural network outputs. Generally, the mean square error function (MSE) has been used for the calculation of the error level. In order to reach the best weight values, some parameters are tuned up such as hidden layer number, hidden neuron number, learning algorithm, activation function and/or kind of neural network in the training stage.

In this study to get the best values, one hidden layer with $4, 7$ and $10$ neurons, Levenberg-Marquardt learning algorithm, tangent hyperbolic activation function and multi-layer feed-forward neural network have been used. By using final weights values, the comparison has been made between neural network outputs and the desired values. In the present study, the inputs were $g_l$, $M_{Z_l}$ and energy for the estimation of the cross-section. Of the total $412$ data available for $380\,GeV$, $1.5\,TeV$ and $3\,TeV$ energy values, $310$ were used in the training stage and $102$ were used in the test stage. During the training of the ANN, no other data sharing was performed. However, the results obtained are given separately in Section 4 according to different energy and $g_l$ values. The range of activation function is (-1; 1) for the hyperbolic tangent of the hidden layer. Therefore, it can be said that it can potentially be difficult to train cases without normalizing or softening the data. Also generally in the method, the data is normalized or smoothed in order to speed up the learning process and increasing the learning rate. In case of data are always positive and their scales vary drastically, one simple way is to use the logarithm transformation of the data. Thus, we have taken the logarithm of the output values in the all calculations.

## 3. Production of the $Z_l$ boson at the CLIC energy stages

The CLIC has been revised with updated center of mass energies, $380\ GeV$, $1,5\ TeV$, $3\ TeV$, and key parameters for these energy stages. In the new generation linear colliders as CLIC, it is very difficult to use the total beam energy for the production cross section of the particles obtained as a result of scattering. For this reason, it is necessary to consider the effects of initial state radiation (ISR) and beamstrahlung (BS). To account for all these effects, we used the beam design parameters given in Table 1.

| Parameters | Stage 1 | Stage 2 | Stage 3 |
|---|---|---|---|
| $E(\sqrt{S})\ GeV$ | 380 | 1500 | 3000 |
| $L(10^{34}\ cm^{-2}s^{-1})$ | 1.5 | 3.7 | 6.0 |
| $N(10^9)$ | 5.2 | 3.7 | 3.7 |
| $\sigma_x(nm)$ | 149 | 60 | 40 |
| $\sigma_y(nm)$ | 3 | 1.5 | 1 |
| $\sigma_z(\mu m)$ | 70 | 44 | 44 |

***Table 1.** Key Parameters of the CLIC energy stages.*

Here, $E$ is the center of mass energy, $L$ is the luminosity, $N$ is the number of particles in bunch, $\sigma_x$ and $\sigma_y$ are RMS transverse beam sizes at Interaction Points (IP) and $\sigma_z$ is the RMS bunch length.

After designing the model in section 2, we implement the Lagrangian (2.3) into the CALCHEP Simulation Program [29] for numerical calculations.

Before proceeding to the calculations, it is necessary to determine the parameter space of the model introduced above. In [30,31], the authors get some limits from the sensitive electroweak data on different kinds of $Z'$ bosons.

In our calculations, prefer the limit value from [30] is used:

$$\frac{M_{Z_l}}{g_l} \geq 7 TeV \tag{3.1}$$

As seen in Table 2, the upper bounds of interaction constants for different mass values of $Z'$ obey the (3.1) condition. The mass-to-coupling ratio is crucial for understanding the properties of the leptonic photon and the nature of this possible new interaction.

| $M_{Z_l}(TeV)$ | $g_l$ |
|---|---|
| 0.38 | 0.05 |
| 0.5 | 0.07 |
| 1.0 | 0.14 |
| 1.5 | 0.21 |
| 2.0 | 0.28 |
| 2.5 | 0.35 |
| 3.0 | 0.42 |

**Table 2.** Upper bounds of $g_l$ for different values of $Z_l$ mass

In the all calculations, we have determined the signal process as $e^+e^- \rightarrow \gamma, Z, Z_l \rightarrow \mu^+\mu^-$ and the background process as $e^+e^- \rightarrow \gamma, Z \rightarrow \mu^+\mu^-$. We have chosen this process as the final state containing the $e^+e^-$ pair has a huge background (i.e. due to bhabha scattering). Also, $\tau^+\tau^-$ pair causes complications in the signal due to $\tau$ decays; it is unobservable $\bar{\nu}\nu$ pair in final state.

Figure 1.a, 1.b and 1.c show the cross section versus mass values of the leptonic photon for different coupling constant values at CLIC with center of mass energies of $380\ GeV, 1,5\ TeV\ and\ 3\ TeV$, respectively.

At all three energy stages, the signal appears to be well above the SM background even for small $g_l$ values. This is due to positive interferences between $g$, $Z$ and $Z_l$ for mass values less than the center of mass energy and for equal and larger values, it is due to the interference of $Z_l$ with $g$ and $Z$. In figure 1.c, it is easily seen that the shift of the cross-section peak from the center of mass energy, especially for large values of $g_l$. This shift is due to ISR and BS effects.

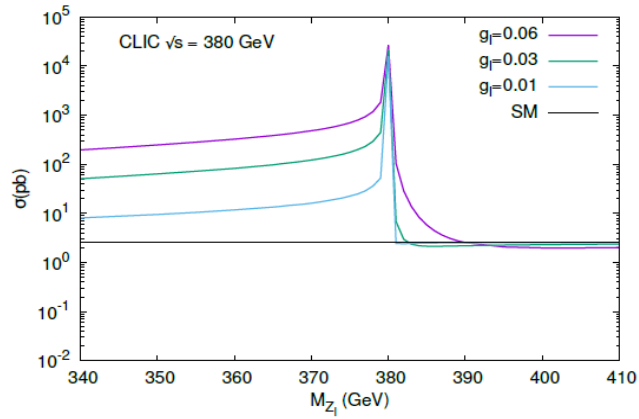

***Figure 1.a*** *Total cross section versus mass values of the leptonic photon for different coupling constant values at CLIC with* $\sqrt{S} = 380\ GeV$.

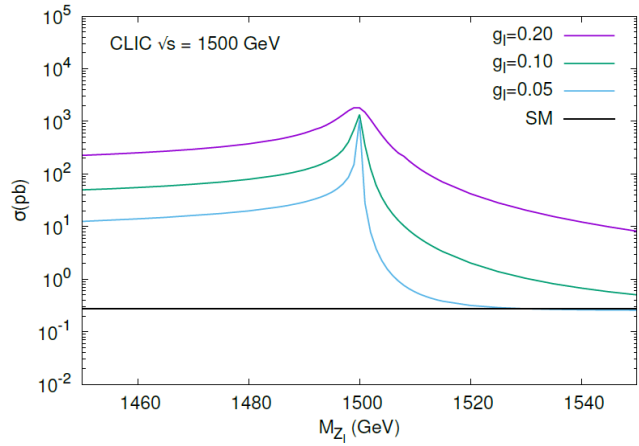

***Figure 1.b*** *Total cross section versus mass values of the leptonic photon for different coupling constant values at CLIC with* $\sqrt{S} = 1,5\ TeV$.

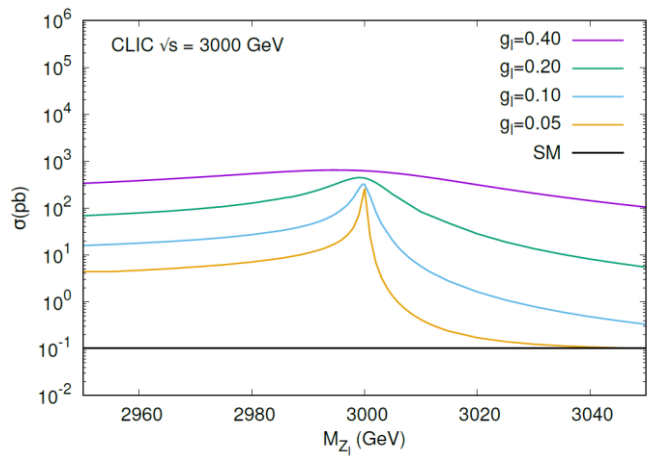

***Figure 1.c*** *Total cross section versus mass values of the leptonic photon for different coupling constant values at CLIC with* $\sqrt{S} = 3\ TeV$.

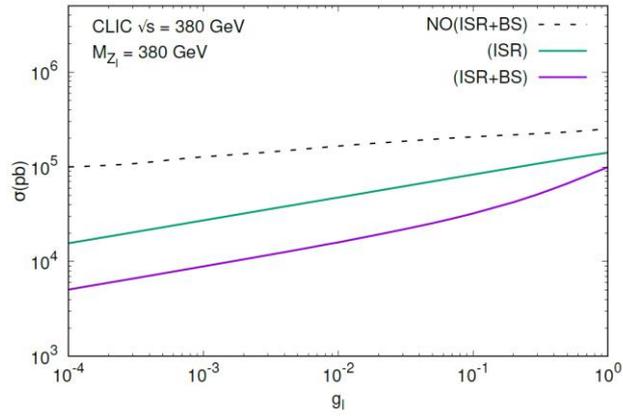

***Figure 2.a*** *For CLIC with $\sqrt{S} = 380\ GeV$ the effects of ISR and BS depending on coupling constant $g_l$*

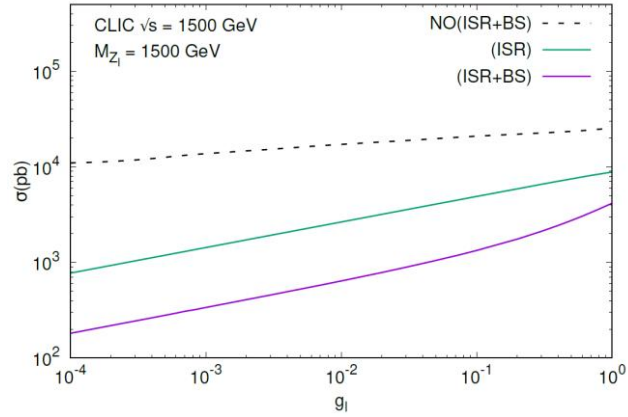

***Figure 2.b*** *For CLIC with $\sqrt{S} = 1,5\ TeV$ the effects of ISR and BS depending on coupling constant $g_l$*

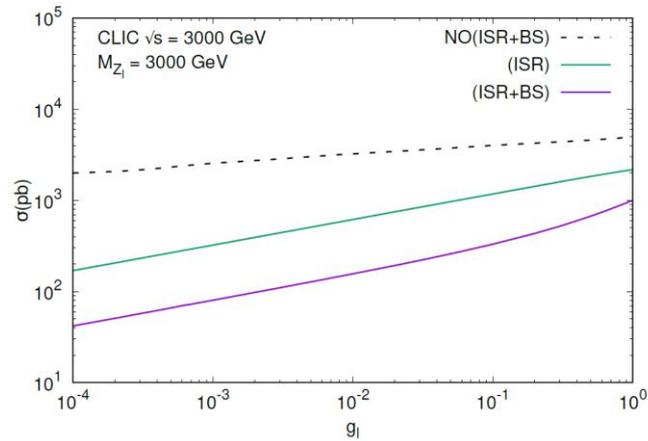

***Figure 2.c*** *For CLIC with $\sqrt{S} = 3\ TeV$ the effects of ISR and BS depending on coupling constant $g_l$*

In figure 2, the cross section versus coupling constant $g_l$ values is plotted to show the effects of ISR and BS together with machine design parameters for CLIC with $\sqrt{S} = 380\ GeV$, $1,5\ TeV$ and $3\ TeV$. It is clear that these effects reduce the corresponding cross-sections at $M_{Z_l} \approx \sqrt{S}$, especially for lower values of $g_l$. All of figure 2 present that ISR and BS effects are more efficient at higher energies: the reduction factor for $g_l = 0,05$ are 7 and 12 at $\sqrt{S} = 380\ GeV$ and $3\ TeV$, respectively.

The following cuts: $|M_{inv}(\mu^+\mu^-) - M_{Z_l}| < 10\ GeV$ and $|\eta_\mu| < 2$ have been used to determine the discovery potential of CLIC at all three energy stages.

Statistical significance $(S)$ is calculated using the following formula:

$$S = \frac{\sigma_{signal} - \sigma_{SM}}{\sqrt{\sigma_{SM}}} \sqrt{L_{int}} \qquad (3.2)$$

where $\sigma_{signal}$ and $\sigma_{SM}$ are the cross-section of signal and background, respectively and $L_{int}$ is the luminosity of the interaction. For CLIC ($\sqrt{S} = 380\ GeV$, $1,5\ TeV$ and $3\ TeV$) $3\sigma$ observations and $5\sigma$ discovery contours against $M_{Z_l}$ and $g_l$ are shown in figure 3. As can be seen from figure 3.a, CLIC with $\sqrt{S} = 380\ GeV$ will give opportunity for massive leptonic photon, $Z_l$, searches in the region from $100\ GeV$ to $400\ GeV$, where the interaction constant is down to $g_l \approx 10^{-3}$. Figures 3.b and 3.c present similar plots for CLIC with $\sqrt{S} = 1,5\ TeV$ and $3\ TeV$, respectively. It is seen from figure 3.c that $Z_l$ could be covered up to $M_{Z_l} = 3\ TeV$ of $g_l \geq 10^{-3}$ for CLIC $\sqrt{S} = 3TeV$.

In figures 4.a-c, the invariant mass distributions of final muons are plotted for signal and SM background at CLIC ($\sqrt{S} = 380\ GeV$, $1,5\ TeV$ and $3\ TeV$). It is clear that the signal is well above the background. As can also be clearly seen from these figures, the condition (3.1) is satisfied.

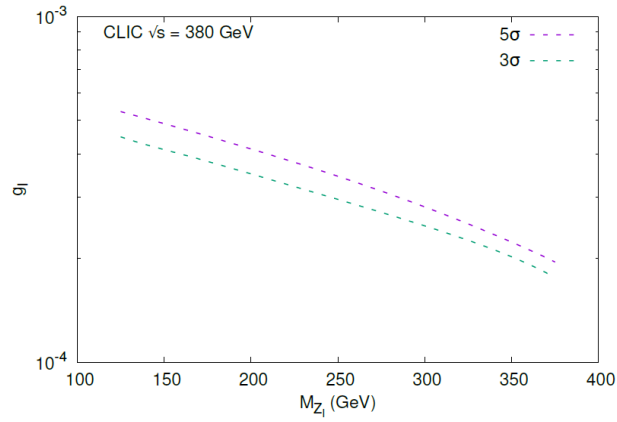

***Figure 3.a*** *At CLIC with $\sqrt{S} = 380\, GeV$ achievable limits (for $3\sigma$ observations and $5\sigma$ discovery) for the mass and coupling parameters.*

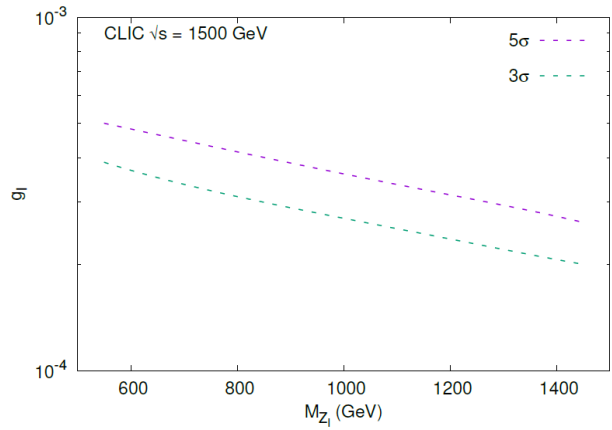

***Figure 3.b*** *At CLIC with $\sqrt{S} = 1{,}5\, TeV$ achievable limits (for $3\sigma$ observations and $5\sigma$ discovery) for the mass and coupling parameters.*

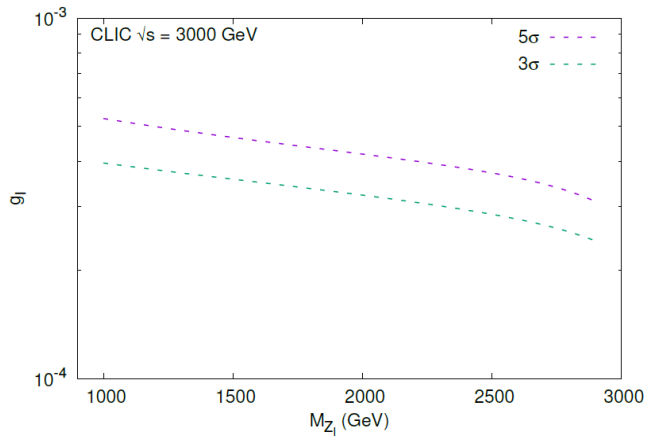

***Figure 3.c*** *At CLIC with $\sqrt{S} = 3\, TeV$ achievable limits (for $3\sigma$ observations and $5\sigma$ discovery) for the mass and coupling parameters.*

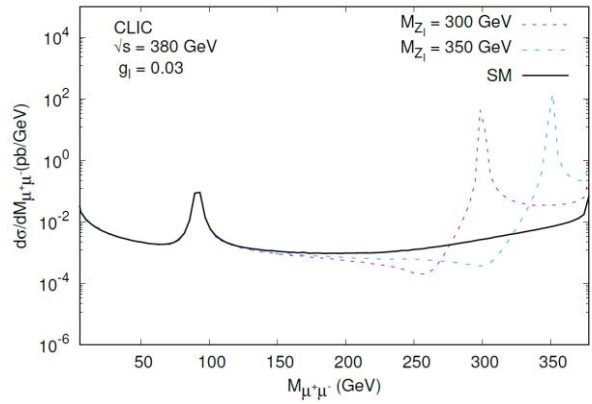

*Figure 4.a* Invariant mass distributions of final muon pairs for signal and SM background at CLIC with $\sqrt{S} = 380\ GeV$ (for two different values of $M_{Z_l}$)

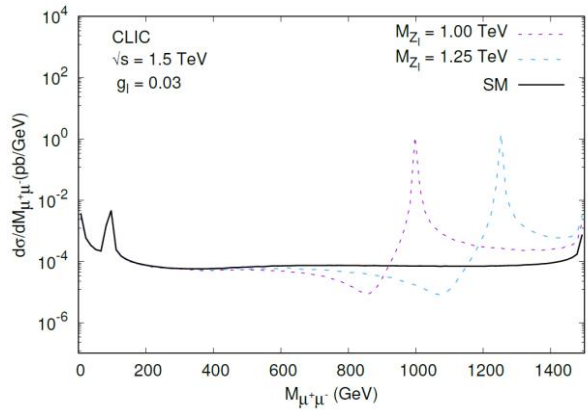

*Figure 4.b* Invariant mass distributions of final muon pairs for signal and SM background at CLIC with $\sqrt{S} = 1,5\ TeV$ (for two different values of $M_{Z_l}$)

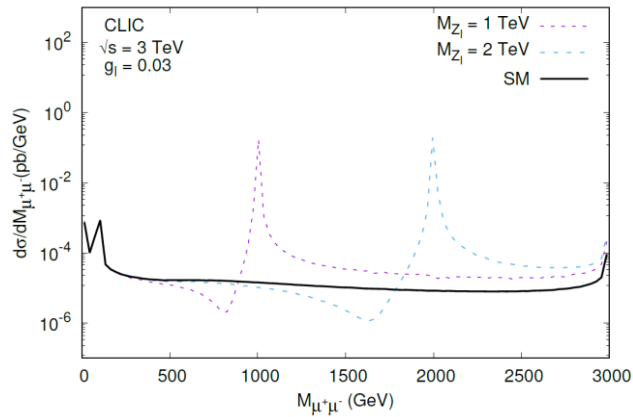

*Figure 4.c* Invariant mass distributions of final muon pairs for signal and SM background at CLIC with $\sqrt{S} = 3\ TeV$ (for two different values of $M_{Z_l}$)

## 4. Predictions by ANN

ANN calculations performed to obtain influence sections were first tested on the training data set. In Figure 5.a-c, the predictions of ANN on this data set are presented separately, according to different energy and gl values, in comparison with theoretical values. When the results given for $380\ GeV$ in the Figure 5.a are examined, it is seen that the ANN structure with $h = 7$ hidden neurons gives more successful results. As can be clearly seen from the graphs, while all three ANN structures are compatible with theoretical data in the high energy region, the ANN with $h = 4$ and $h = 10$ structures are relatively closer to the theoretical values at low energies. However, results compatible with the peak at $380\ GeV$ were obtained with the $h = 7$ ANN structure. When the Figure 5.b is examined, results of $1.5\ TeV$ energy values are given for different $g_l$ values. At this energy value, it is clearly seen that the $h = 7$ ANN structure is successful in capturing the peak in the cross section. Finally, in the Figure 5.c, the prediction results for $3\ TeV$ are presented. The sudden increase in the peak could be achieved with the $h = 7$ ANN structure. In the high and low energy region, it is seen that the ANN structure with $h = 4$ is more successful.

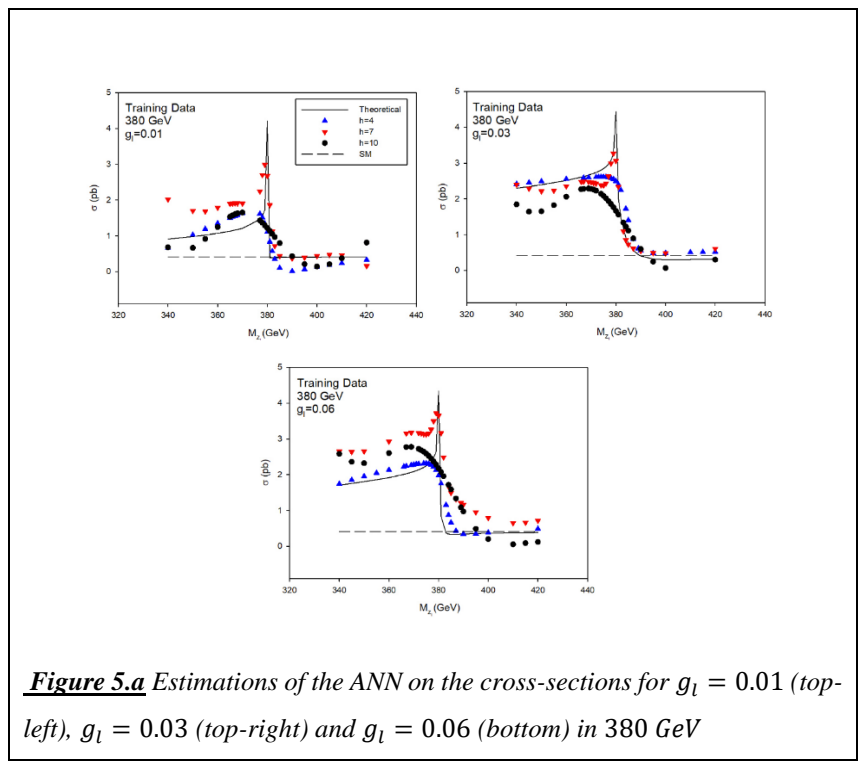

*Figure 5.a* Estimations of the ANN on the cross-sections for $g_l = 0.01$ (top-left), $g_l = 0.03$ (top-right) and $g_l = 0.06$ (bottom) in $380\ GeV$

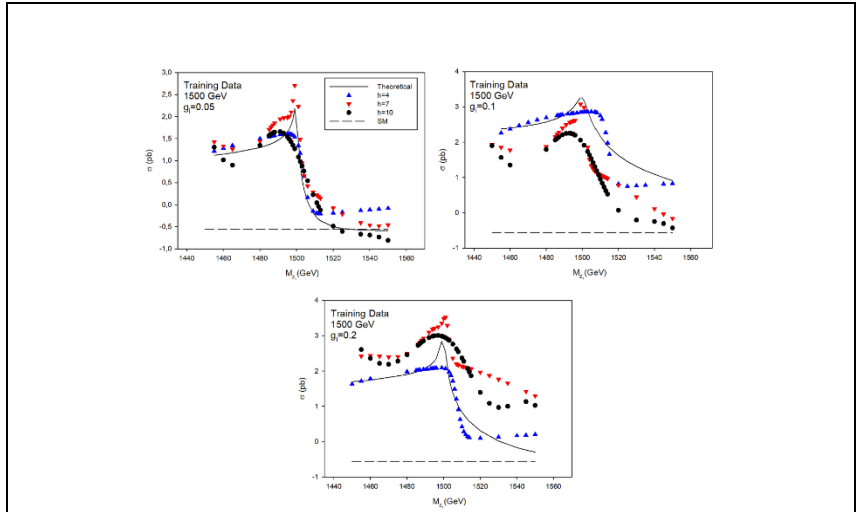

***Figure 5.b*** *Estimations of the ANN on the cross-sections for $g_l = 0.05$ (top-left), $g_l = 0.1$ (top-right) and $g_l = 0.2$ (bottom) in $1.5\ TeV$*

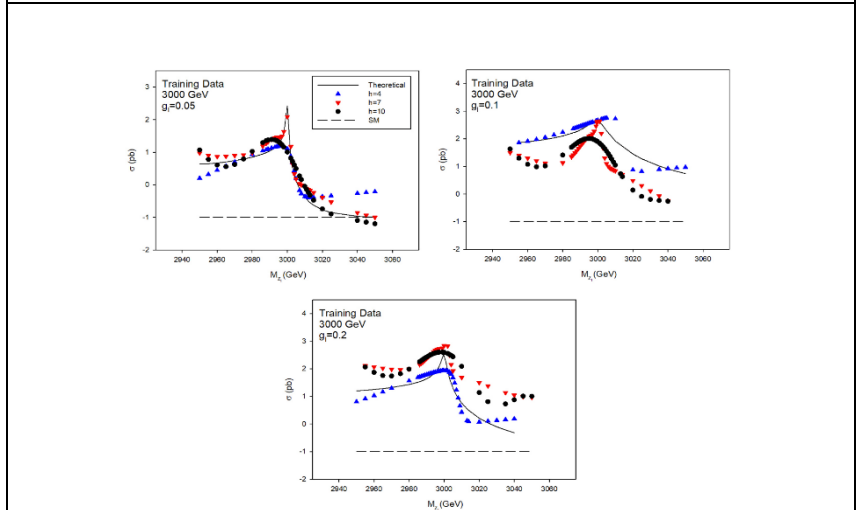

***Figure 5.c*** *Estimations of the ANN on the cross-sections for $g_l = 0.05$ (top-left), $g_l = 0.1$ (top-right) and $g_l = 0.2$ (bottom) in $3\ TeV$*

The prediction results of ANN calculations on test data are given in Figure 6.a-c, in comparison with theoretical results. In the figures, the differences between the results of the calculations for $380\ GeV$, $1.5\ TeV$ and $3\ TeV$ from the theoretical values are shown, respectively. For $380\ GeV$, the $h = 10$ structure in the case of $g_l = 0.01$, the $h = 4$ structure in the case of $g_l = 0.03$ and the $h = 4$ structure in the case of $g_l = 0.06$ generally made more successful predictions. When the graphs for $1.5\ TeV$ are examined, it is clear that the $h = 4$ structure for $g_l = 0.05$ and $g_l = 0.1$ and the $h = 10$ structure for $g_l = 0.2$ give more successful results compared to the others. For the 3 $TeV$ energy value, ANN with $h = 4$ structure for $g_l = 0.05$ and ANN with $h = 10$ structure for $g_l = 0.1$ and $g_l = 0.2$ made better predictions.

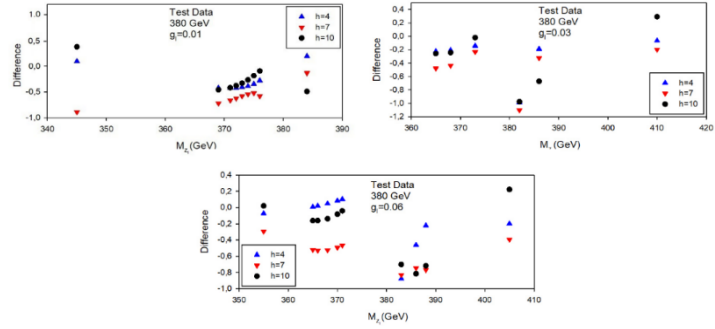

**_Figure 6.a_** Test dataset predictions of the ANN on the cross-sections for $g_l = 0.01$ (top-left), $g_l = 0.03$ (top-right) and $g_l = 0.06$ (bottom) in 380 $GeV$

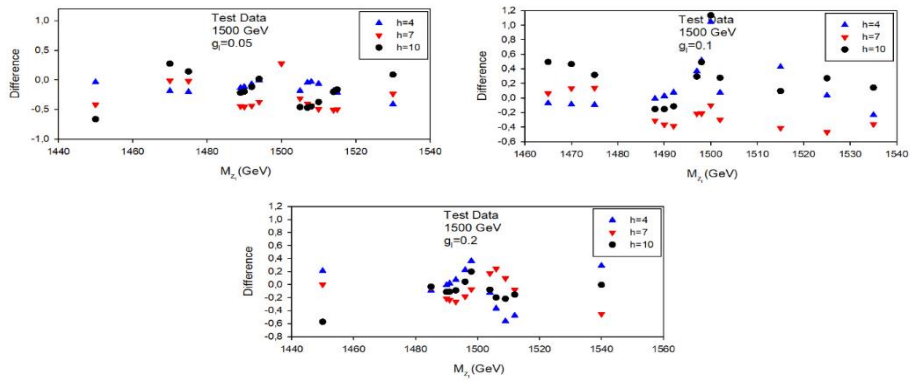

**_Figure 6.b_** Test dataset predictions of the ANN on the cross-sections for $g_l = 0.05$ (top-left), $g_l = 0.1$ (top-right) and $g_l = 0.2$ (bottom) in 1.5 $TeV$

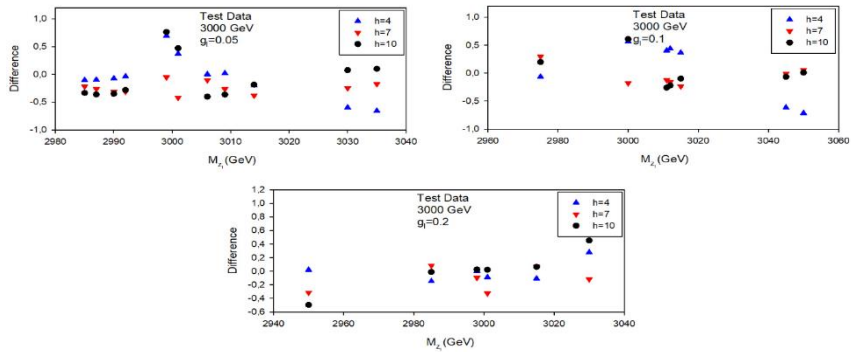

**_Figure 6.c_** Test dataset predictions of the ANN on the cross-sections for $g_l = 0.05$ (top-left), $g_l = 0.1$ (top-right) and $g_l = 0.2$ (bottom) in 3 $TeV$

In Figure 7, graphs of ANN predictions on the training and test data sets compared to theoretical results are presented separately. When the scatter plot of the training data set is examined, it is seen that the distribution for $h = 10$ is narrower, while the others are relatively more widespread. When the distributions of the test data set were examined, it was seen that the $h = 4$ ANN structure showed a less widespread distribution.

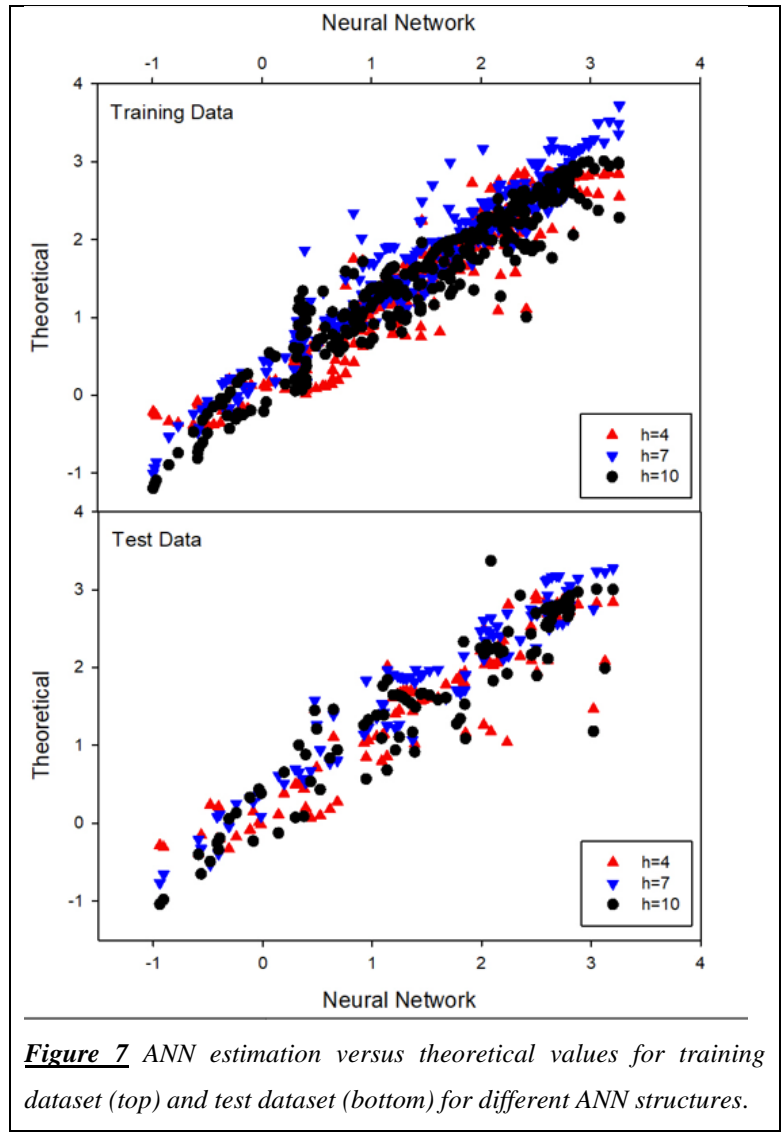

*Figure 7* ANN estimation versus theoretical values for training dataset (top) and test dataset (bottom) for different ANN structures.

## 5. Conclusions

In this paper, we have shown that massive leptonic vector boson, $Z_l$, with masses up to the center of mass energy can be observed using the process $e^+e^- \rightarrow \mu^+\mu^-$ at the new generation linear collider CLIC with updated center of mass energies ($\sqrt{S} = 380\ GeV, 1,5\ TeV$ and $3\ TeV$) and machine design parameters, provided that $g_l \geq 10^{-3}$. This study presents hypotheses to the literature regarding some fundamental properties, such as mass and coupling constant, of a possible new massive vector boson, $Z_l$. From the calculations and the results, it can be concluded that ISR and BS will have a significant impact on the investigation of $Z_l$ at CLIC. At higher center of mass energy such as CLIC with $\sqrt{S} = 3\,TeV$, this impact becomes more important. When the results obtained in ANN calculations were examined, it was seen that the ANN structure with 10 hidden layer neurons ($h = 10$) was more successful in the training phase, and the structure with 4 hidden layer neurons ($h = 4$) was more successful in the testing phase. However, the ANN structure with 7 hidden layer neurons ($h = 7$) was more successful in more accurately predicting the peaks corresponding to the sudden increase in the cross sections. As a result, it is concluded that ANN can be used as an alternative tool in estimating influence sections.